# THE ACCELERATION AND STORAGE OF RADIOACTIVE IONS FOR A BETA-BEAM FACILITY

Mats Lindroos
*AB-division, CERN, Geneva, Switzerland*

The beta-beam working group
*http://cern.ch/beta-beam*

The term beta-beam has been coined for the production of a pure beam of electron neutrinos or their antiparticles through the decay of radioactive ions circulating in a storage ring. This concept requires radioactive ions to be accelerated to as high Lorentz $\gamma$ as 150. The neutrino source itself consists of a storage ring for this energy range, with long straight sections in line with the experiment(s). Such a decay ring does not exist at CERN today, nor does a high-intensity proton source for the production of the radioactive ions. Nevertheless, the existing CERN accelerator infrastructure could be used as this would still represent an important saving for a beta-beam facility.

## 1. Introduction

The proposed beta-beam facility [1] can be divided in two parts, the low energy part stretching up to 100 MeV/u and the high energy part for the acceleration to higher energies and for the stacking process in the decay ring (see figure 1). The division is logical as the low energy part corresponds to the requirements for a radioactive beam factory of ISOL type as proposed and promoted by the Nuclear Physics community. The high energy part, which serves the neutrino physics community, would only be one of several clients of such a radioactive beam facility and would consequently share the cost and use of the low energy part with other physics communities.

The radioactive ions ($^6$He and $^{18}$Ne) will be produced in an ISOL system using the proposed Superconducting Proton Linac (SPL) as a driver. They will be ionised, bunched and accelerated with a linac to approximately 100 MeV/u. From the linac they will be injected into a storage ring and ejected to a fast cycling synchrotron for acceleration to 300 MeV/u. The ions will be extracted into the existing Proton Synchrotron (PS) and accelerated to PS top energy in several bunches. Following this they will be transferred to the Super Proton Synchrotron (SPS) and accelerated to top energy. Finally, they will be transferred to the new decay ring where they will be merged with the already circulating bunch through a longitudinal stacking procedure. While this might appear straightforward, several bottlenecks exist in this process, not least the bunching at low energy, space charge limitations in the PS and the SPS, decay losses in the acceleration chain and the longitudinal stacking procedure at high energy in the decay ring. In this paper we will discuss the problems encountered and some possible solutions that have been proposed.

## 2. Ion production

The flux of neutrinos in the decay ring will be determined by the half life at a given Lorentz γ of the radioactive ions and the equilibrium intensity stored. The flux at the detector will further depend on the average energy of the neutrinos at rest as this will determine the focusing of the neutrino beam. A further constraint is set by the decay losses in the accelerator chain that will increase with shorter half lives and possible decay products that could create long lived contamination in the low energy part. All these constrains together points towards two isotopes of particular interest, $^6$He giving a beam of anti-ν and $^{18}$Ne for ν [2]. Both can be produced in large quantities through the so called ISOL method. The helium isotope is best produced in a beryllium target using a very intense primary proton beam impinging on a so called neutron converter (fission) and the Neon isotope through spallation of a MgO target with a less intense 2 GeV proton beam hitting the target material directly. It is possible to produce a factor of ten more helium isotopes than neon isotopes thanks to the converter technology.

## 3. Ionization, bunching and pre-acceleration

The isotope can be transported from the ISOL target in gas form given the fact that the chosen elements are noble gases. Alternatively a high efficiency (for noble gases) mono charge ECR source [3,4] can be used close to the target and the singly charged ions can be transported using classical beam transport. In either case the beam has subsequently to be bunched for the further acceleration and injection into synchrotrons. Efficient bunching (<20 microseconds) and full stripping of a high intensity beam can be achieved using a high frequency (60 GHz) ECR source [5]. While such a system does not exist today theoretical calculations show that it can be constructed. Furthermore, the necessary components will soon be available and the required RF generator is available off the shelf and could be assembled for a first feasibility test in the close future.

Acceleration of high intensity radioactive beams to 100 MeV/u has already been studied within the EU financed RTD study EURISOL [6] using a linac. This study is planned to be pursued in a design study within the EU 6$^{th}$ framework programme.

## *4. Acceleration*

The existing CERN accelerator infrastructure can handle high intensity ion beams but space charge constrains at injection into the first major CERN synchrotron, the PS, would require an injection energy of at least 300 MeV/u for the intensities required for the beta-beam. Furthermore, the linac beam will indeed be bunched (from the ECR source) but will need further shortening to fit the necessary intensity into the PS. A first stage of additional bunching and fast acceleration is pictured to take place in a combined storage ring and rapid cycling synchrotron. The most promising scenario is injection by classical multi-turn injection of the linac beam into the storage ring followed by fast acceleration and transfer to the PS. The ion bunches will wait in the PS until all buckets (8-16 bunches) are filled for acceleration to higher energies. The top energy of the PS will depend on the mass charge ratio of the ions and will be slightly different for Neon and Helium. The transfer to the next CERN synchrotron, the SPS, will encounter space charge limitations [2], which will require that the transverse emittance is increased possibly by multiple passages through a foil. The SPS has to be fitted with a new 40 MHz RF system to handle these long bunches but transfer to the existing 200 MHz system can take place at the transition energy, something that will reduce the costs for the necessary modifications [2].

## *5. Stacking*

The energy of the ν and anti-ν beam from the beta-beam facility will be in the range of the atmospheric neutrinos so the ions have to be kept in short bunches in the decay ring to permit a efficient back-ground suppression in the detector. A target value of in total four bunches not longer than 10 ns each have been chosen for the conceptual design. The intensity of the individual bunches can be increased through stacking of neon and helium ions. Stacking has been used for many particle types in different applications and consists of a regular re-fill of the bunch before the already present particles ceased being useful. Many recent stacking methods use some form of beam cooling to reduce the phase space density and make room for more particles within a given transverse and longitudinal emittance. Cooling times for heavy light ions at high energies using classical methods such as electron cooling and stochastic cooling are long and can not easily be reduced. A new scheme of longitudinal stacking has been proposed for the beta beam [2] using a combination of bunch rotation and asymmetric bunch merging. The freshly injected ion bunches (as many as permanently circulating bunches) are initially placed on a dispersion orbit to avoid that the injection elements, septa and kicker magnets, influence the circulating beam. In a second stage the fresh beam is rotated in longitudinal phase space which means that they lose or gain energy in such a way that they are physically moved to the orbit of the residing bunches. This is achieved by capturing the residing bunches and the injected bunches in a common large RF bucket in which the bunches can rotate ending up on the same orbit. In the next stage the RF bucket is manipulated by an additional RF system so that the injected bunch is moved and mixed with an equal proportion of the centre of the residing bunch. This is a common beam manipulation in synchrotrons named bunch merging or splitting, but in this case it is modified so that it only affects the actual populated part of phase space avoiding merging a large proportion of empty space from the smaller injected bunch with the residing bunch. The fact that only the central part of the residing bunch is affected will result in a net increase in the core intensity pushing the surrounding "older" ions out towards the bucket separatrix where eventually the "oldest" ions will be lost. The latter part of this operation - the asymmetric bunch merging - has recently been tested in the CERN PS with success [7]. Empty phase phase space from an imaginary adjacent smaller bucket was merged with a high intensity proton beam (see figure 2). The PS RF system could be modified to permit full longitudinal beam control of the process yielding a high degree of efficiency.

## *6. Intensity*

The maximum number of particles that can be stored in the decay ring just after a completed stacking cycle is, assuming no beam losses,

$$N_{tot} = N_{bunch} \frac{1}{1 - 2^{-\frac{T}{(\gamma T_{half})}}} \qquad (1)$$

where $N_{bunch}$ is the number of particles with half-life $T_{half}$ at Lorentz $\gamma$ injected into the ring every T seconds [1]. The decisive number for the neutrino flux in the detector is the average number of ions in the decay ring between to stacking cycle which can be calculated from (1) as

$$N_{average} = \frac{N_{bunch} \gamma T_{half}}{T \operatorname{Ln}(2)} \qquad (2)$$

The experience at CERN is that the losses through the accelerator chain will be less than 50% excluding decay losses. Taking the given production rates after the ECR source, allowing for decay in the accelerator chain we arrive at the values given in table 1 for the intensities stored for our conceptual design. Note that the $\gamma$ for the Neon ions is only 70 compared to 150 for the helium ions. Early simulations showed that this was a reasonable approach.

| Machine | Ions extracted | Batches | Loss power | Losses/length |
|---|---|---|---|---|
| Source + Cyclotron | $2 \cdot 10^{13}$ ions/s | 52.5 ms | N/A | N/A |
| Storage Ring | $1.02 \cdot 10^{12}$ | 1 | 2.95 W | 19 mW/m |
| Fast Cycling Synchrotron | $1.00 \cdot 10^{12}$ | 16 | 7.42 W | 47 mW/m |
| PS | $1.01 \cdot 10^{13}$ | 1 | 765 W | 1.2 W/m |
| SPS | $0.95 \cdot 10^{13}$ | ∞ | 3.63 kW | 0.41 W/m |
| Decay Ring ($N_{tot}$) | $2.02 \cdot 10^{14}$ | N/A | 157 kW | 27.96 W/m |

| Machine | Ions extracted | Batches | Loss power | Losses/length |
|---|---|---|---|---|
| Source + Cyclotron | $8\ 10^{11}$ ions/s | 52.5 ms | N/A | N/A |
| Storage Ring | $4.14\ 10^{10}$ | 1 | 0.18 W | 1.1 mW/m |
| Fast Cycling Synchrotron | $4.09\ 10^{10}$ | 16 | 0.46 W | 2.9 mW/m |
| PS | $5.19\ 10^{11}$ | 1 | 56.4 W | 90 mW/m |
| SPS | $4.90\ 10^{11}$ | $\infty$ | 277 W | 32 mW/m |
| Decay Ring (Ntot) | $9.11\ 10^{12}$ | N/A | 10.6 kW | 1.88 W/m |

Table 1: Intensities and average loss power for the $^6$He (top) and $^{18}$Ne (bottom) beam, assuming a 16 Hz fast cycling synchrotron and 8 s SPS cycle time. Only beta-decay losses are taken into account. To get an approximate number for Ntot with losses taken into account, the number of ions in the decay ring should be multiplied with a factor 0.5.

## *7. Decay losses*

The main difference between the acceleration to high energies of stable ions and radioactive ions are the losses induced by the radioactive decay during the acceleration process. The isotopes proposed for the beta-beam conceptual design have been chosen so that no long-lived activity is left to contaminate the low energy part. At high energies the ions are likely to break up and it has been shown that the induced activity is well approximated by the simultaneous loss of the same number of protons as the total A of the studied ion type [8]. A first simulation of losses in the decay ring for the conceptual design [9] yields that the dose rate of the arcs is limited to 2.5 mSv/h after 30 days of operation and 1 day of cooling. Furthermore, the induced radioactivity in the ground water will have no major impact on public safety. It has also been confirmed that the total loss in the accelerator chain will be beyond the nominal 1 W/m for hands on maintenance except in the PS [10], something that will require further thought, see table 1.

## *8. Possible improvements*

A number of more speculative improvements to the conceptual design have recently been proposed [11], they will all require much more work to reach the same maturity as the parameters in the conceptual design. However, it is worth presenting them here to permit a larger community to participate in the evaluation process of their feasibility.

The decay losses in the PS can be dramatically reduced by the construction of a new fast cycling synchrotron. The synchrotron should also have a higher maximum magnetic rigidity permitting a reduction of space charge constrains at PS-SPS transfer. An increase of the Neon production is highly wished for and a possible design with multiple targets in sequence making use of the same proton beam has been proposed. It is imagined that the Neon production could be increased with as much as factor of three using this concept [12]. A very important improvement of the ν and anti-ν flux could be achieved by storing Neon and Helium ions simultaneously in the decay ring [13]. This is possible even if the mass to charge ratio is different for these two types of ions if the required difference in mean bending radius is small or if an artificial total orbit length is added through an insertion for one particle type. In the latter case it is probably necessary to consider the use of two independent RF systems for the two particle types. That would also simplify the stacking procedure as the two ion types would be longitudinally independent. The difference in mean bending radius ρ is, assuming a fixed magnetic rigidity, $B\rho_{dipole}$, set by the magnetic field B and the bending radius of the dipoles $\rho_{dipole}$ in the ring:

$$\Delta\rho = \rho_{Neon} - \rho_{Helium} = \frac{\beta_{Neon}}{\beta_{Helium}}\left(\frac{SS_{Helium}}{\pi} + \rho_{Helium}\right) - \frac{SS_{Neon}}{\pi} - \rho_{Helium} \quad (3)$$

where SS is the length of the straight section for each ion type (with insertion). This will impose a fixed relationship between the Lorentz γ of the two ion types

$$\gamma_{Neon} = \frac{A_{Helium}}{q_{Helium}} \bigg/ \frac{A_{Neon}}{q_{Neon}} \times \gamma_{Helium} \quad (4)$$

The acceleration of two ion types for simultaneous use in the decay ring will require separate acceleration cycles, something that necessarily will lengthen the interval between stacking in the decay ring which will reduce the maximum number of ions stored and require special RF manipulations (e.g independent RF systems) to avoid disturbing the "wrong" ions during stacking. An increased longitudinal acceptance can compensate for most of the losses. Evidently, the maximum stored intensity can also be increased by a shorter total accelerator cycle. The actual magnetic cycle time for the SPS, which determines the length of the overall cycle can probably be decreased if the beta-beam was the only user of the SPS. However, the beta-beam facility is imagined to be taken into operation while LHC operation, CNGS operation and possibly new fixed target physics still are part of the CERN physics programme. Still, an optimization of the machine use can probably yield a shorter average cycling time, in particular if the required maximum γ for ions can be reduced.

Taking all this improvements into account a new simulation of the physics reach for this more speculative scenario has been performed [14]. While the earlier simulations [15] based on the conceptual design are already showing an impressive physics reach and clearly motivates a continued study of a beta-beam facility, the new simulation results show that the here discussed upgrade options are well worth pursuing. The target values from the latest simulation [14] for γ and intensity in the decay ring with both ion types present are for helium at γ=60 a total of $4 \times 10^{13}$ ions in the ring and for neon at γ=100 a total of $2 \times 10^{13}$ in the ring.

It also been proposed that experiments relevant for astrophysics and nuclear structure could be performed with a low energy beta-beam [16]. The advantage compared to static neutrino sources, such as nuclear power plants and spallation sources are the possibility to choose either a ν beam or an anti-ν beam and the fact that the neutrinos are delivered as a beam with a preferred direction. There are two possibilities, either experiments in a close detector at the here proposed beta-beam facility or a dedicated low energy ring with possibly other more long lived isotopes and an almost continues injection to produce a very intense ν or anti-ν beam.

## *9. Conclusions*

The conceptual design for the beta-beam is based on realistic estimates and conservative extrapolations. The only point that raises some concern is the losses in the PS which would make it difficult to continue hands on maintenance for at least parts of the machine. It is probably necessary to at least envisage an upgrade of the cycling time to reduce these losses and increase the overall efficiency of the accelerator chain.

The more speculative scenario with increased Neon production and simultaneous storage of both ion types requires much more work before it can be set on the same firm ground as the conceptual design. Still, it is clearly well worth pursuing.

The low energy beta-beam opens new fields of physics for a beta-beam facility. A detailed analysis of the requirements on the flux is in progress together with first estimations for the design parameters of the associated low energy decay ring.

[16] C. Volpe, What about a beta-beam facility for low energy neutrinos?, hep-ph/0303222, Submitted to Phys. Rev. Lett.

## *Figures*

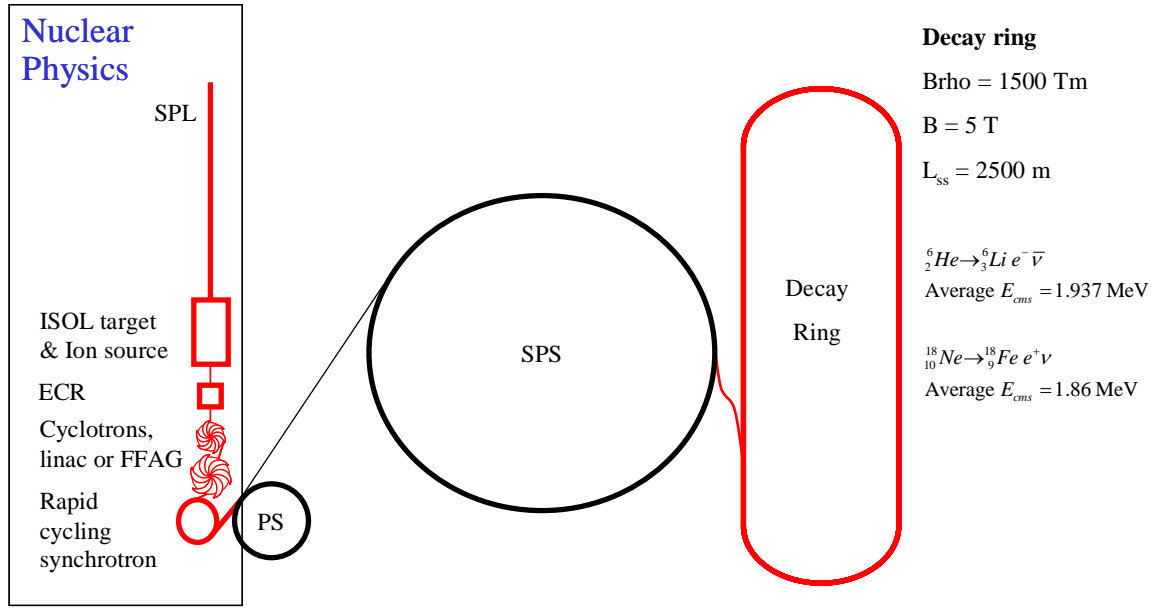

Figure 1: The CERN conceptual design

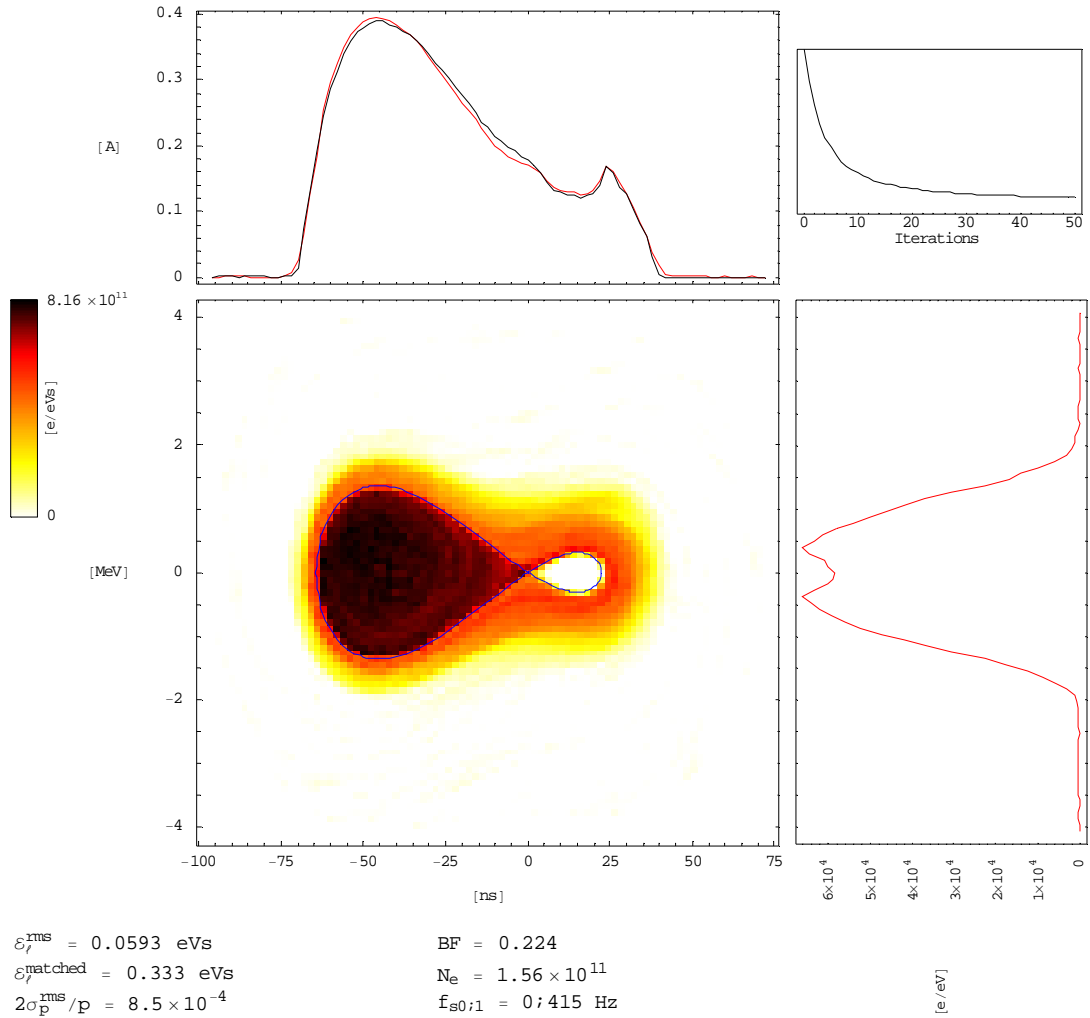

Figure 2. Tomographic reconstruction of asymmetric bunch merging in the CERN PS. A high intensity proton bunch is asymmetrically merged with large efficiency with smaller empty bucket [7].